\documentstyle[cjaa209,twoside,epsf]{article}
\input{psfig.sty}                               

\begin{document}

\title{Unveiling the nature of {\it INTEGRAL} sources \\ through optical 
spectroscopy}

\author{Nicola Masetti\mailto{masetti@bo.iasf.cnr.it}}

\inst{INAF -- Istituto di Astrofisica Spaziale e Fisica Cosmica, Sezione
di Bologna, via Gobetti 101, I-40129 Bologna (Italy)}

\email{masetti@bo.iasf.cnr.it}

\markboth{N. Masetti}{The nature of {\it INTEGRAL} sources via optical 
spectroscopy}

\pagestyle{myheadings}

\vspace{-.2cm}
\date{Received August 8, 2005; accepted April 20, 2006}

\baselineskip=12pt

\vspace{-.3cm}
\begin{abstract}
\baselineskip=12pt
Since its launch on October 2002 the {\it INTEGRAL} satellite is 
performing an deep survey of the hard X--ray sky with unprecedented 
sensitivity and positional accuracy. This allowed pinpointing, through 
positional cross-correlation with catalogs at longer wavelengths, possible 
optical/near-infrared candidates for the hard X--ray sources of still 
unknown nature. In this presentation I will describe this work as well as 
the observational activities aimed at determining, through optical 
spectroscopy, the nature of the unknown {\it INTEGRAL} sources, along with 
the main results of this search. Future prospects about this 
identification program will also be illustrated.

\vspace{-.2cm}
\end{abstract}

%
%
\vspace{-.9cm}
\section{Introduction}    

\vspace{-.3cm}
One of the main aims of satellites observing in the X--ray band is 
to obtain all-sky maps of celestial high-energy emission. This allows 
obtaining information on sky distribution and characteristics of X--ray 
objects. In the past years, several surveys were performed by various 
spacecraft such as, e.g., {\it HEAO-1} (13--180 keV, Levine et al. 1984), 
{\it ROSAT} (0.1--2.4 keV, Voges et al. 1999) and
BATSE onboard {\it Compton-GRO} (25--160 keV, Shaw et al. 2004).

These surveys were mostly devoted to all-sky scannings, with particular
attention to the Galactic Plane. In particular, the survey performed by
{\it ROSAT} (Voges et al. 1999) allowed pinpointing, with a precision of
few arcsecs, sources emitting in the soft X--ray band (below 2 keV).
However, it could not detect heavily absorbed sources, in particular those
located along the Galactic Plane, as the high amount of neutral hydrogen 
present in this strip of the sky severely hampers observations in this 
band. Likewise, the main drawbacks of the past hard X--ray surveys were 
the scarce positional accuracy (allowing source error boxes with radii not 
smaller than some degrees) and/or the low survey sensitivity ($\sim$30 
mCrab).

In this sense, {\it INTEGRAL} (Winkler et al. 2003) produced a 
breakthorugh in the all-sky
mapping of hard X--ray sources in terms of both sensitivity and positional
accuracy. Indeed, thanks to the capabilities of the IBIS instrument
(Ubertini et al. 2003), {\it INTEGRAL} is able to detect hard X--ray
sources at the mCrab level with a typical localization accuracy of 2-3$'$:
this has made it possible, for the first time, to resolve crowded regions
such as the Galactic Centre and the spiral arms, and to discover many new
hard X--ray extragalactic objects beyond the Galactic Plane of the Milky
Way, in the so-called `Zone of Avoidance', where (as remarked above) the 
massive presence of interstellar neutral hydrogen disfavours observations 
in soft X--rays.

\vspace{-.4cm}
\section{The 1$^{\rm st}$ IBIS/{\it INTEGRAL} survey}

\vspace{-.3cm}
In its first year, during the 1$^{\rm st}$ survey of the Galactic Plane,
the IBIS instrument onboard {\it INTEGRAL} has detected 123 sources
between 20 and 100 keV (Bird et al. 2004): within this sample of hard
X--ray emitting objects, 53 low-mass X--ray binaries (LMXBs) and 23
high-mass X--ray binaries (MXRBs), 5 Active Galactic Nuclei (AGNs) and a
handful of other objects such as pulsars, supernova remnants and
cataclysmic variables (CVs) were identified. The remaining objects 
(28, or about 23$\%$ of the sample) have no obvious conterparts at other
wavelengths and therefore cannot yet be associated with any known class of
high-energy emitting objects. Only for a tiny fraction of these sources
follow-up observations at X--ray energies as well as in the 
optical/near-infrared (NIR) wavebands have been carried out so far (Rodriguez
2005). Although the cross-correlation with catalogues or surveys at other
wavelengths (especially soft X--rays, optical and radio) is of invaluable
help in pinpointing the putative optical candidates, only accurate optical
spectroscopy can confirm the association and reveal the nature of the
object.

Most of these unidentified sources are believed to be X--ray binary
systems, where one of the two members is either a black hole or a neutron
star (NS). There is however the possibility that some of them could be AGN
similar to those already detected (Bassani et al. 2004). However, since
all these objects are hard X--ray selected and poorly known at other
wavebands, there are possibilities that we are dealing with peculiar 
sources (e.g. Filliatre \& Chaty 2004).

\vspace{-.4cm}
\section{Correlations with catalogs at other wavelengths}

\vspace{-.3cm}
Stephen et al. (2005) studied the positional correlation between the 
sources of the 1$^{\rm st}$ IBIS survey and those in the larger 
{\it ROSAT} bright source catalogue (Voges et al. 1999). It was found 
that, assuming a (conservative) 3$'$ radius for all {\it INTEGRAL} error 
boxes, there are 75 {\it INTEGRAL}/{\it ROSAT} positional associations.
On purely statistical grounds, Stephen et al. (2005) moreover demonstrated 
that, if these catalogs were uncorrelated, the number of expected chance 
associations instead would have been 0.35, that is, a factor of about 200 
times less than the actual number of associations found. This strongly
indicates that a {\it ROSAT} source within an {\it INTEGRAL} error box 
corresponds to the soft X--ray counterpart of the hard X--ray object 
detected with IBIS; this fact allows one to reduce the error box of the 
X--ray source to few arcsecs, facilitating the searches of the optical 
counterpart.

Of the 75 {\it ROSAT} sources found associated with {\it INTEGRAL} 
detections, only 66 have an identified counterpart at longer wavelengths, 
in such a way that their nature is known (Stephen et al. 2005);  
this leaves 9 {\it INTEGRAL} objects not yet optically identified but
for which, statistically, the corresponding {\it ROSAT} source should be 
the counterpart at softer X--rays.

Despite the strong correlation found between the two catalogues, a
significant fraction ($\sim$40\%) of the {\it INTEGRAL} sources have no
association with a {\it ROSAT} bright survey object. This, as said in
Sect. 1, may be due to strong absorption preventing detection in soft 
X--rays.

A similar, although less tight, correlation seems to be present (Stephen
et al., in preparation) between the IBIS survey and radio catalogs such as
the NVSS (Condon et al. 1998); thus, the presence of a radio object within
the IBIS error box can be seen as an indication of an association between 
the radio and the {\it INTEGRAL} sources (see, e.g., Masetti et al. 2004).

\vspace{-.4cm}
\section{Optical/NIR identifications of unknown {\it INTEGRAL} sources}

\vspace{-.3cm}
The first identification of the nature of an unknown source discovered
with {\it INTEGRAL} was that of IGR J16138$-$4848, an object located in
the Norma Arm of the Galaxy. Thanks to the refined (error radius: 4$''$)
{\it XMM-Newton} soft X--ray position, Filliatre \& Chaty (2004)
pinpointed the using optical and NIR photometry. Then, through the use of
NIR spectroscopy, these authors identified this source as an extremely
reddened (with optical $V$-band absorption $A_V$ = 17.4 magnitudes) MXRB,
located at a distance between 0.9 and 6.2 kpc, and composed of a 
supergiant B[e] star losing mass onto a compact object, most likely a NS.

Motivated by the findings illustrated in Sect. 3, and in parallel with the
above successful identification of the nature of IGR J16138$-$4848, we
underwent a sample selection in order to pick out possible optical
candidates over which optical spectroscopy could be done in order to 
unveil their nature.

We performed the sample selection following three steps: (i) we positionally
correlated the IBIS survey with the {\it ROSAT} (X--ray) and NVSS (radio)
catalogues in order to substantially reduce the error box size; (ii)
within these reduced error boxes we searched for putative optical/NIR
counterparts on DSS-II-Red (optical) and 2MASS (NIR) surveys; (iii) we
selected cases with few (3 or less) relatively bright optical/NIR
candidates in the {\it ROSAT} and/or NVSS error boxes on which optical
spectroscopy could be performed. We then started our campaign with three
objects observable (in terms of both declination and optical brightness)
with the 1.5-metre `Cassini' telescope of the Astronomical Observatory of
Bologna in Loiano (Italy): IGR J17303$-$0601 ({\it the Good}), IGR
J18027$-$1455 ({\it the Bad}) and IGR J21247+5058 ({\it the Ugly}). For
details, the reader is referred to the paper by Masetti et al. (2004a).

\begin{figure}[t!]
\hspace{2cm}
\psfig{file=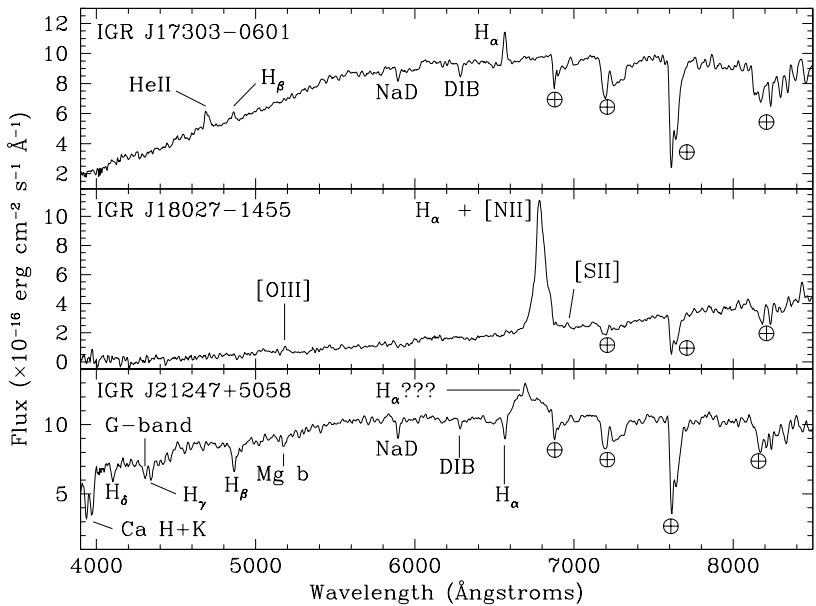,width=11cm,angle=270}
\vspace{-.3cm}
   \caption{Average optical spectra, not corrected for Galactic or
intrinsic absorption, of the optical counterparts to IGR
J17303$-$0601 (upper panel), IGR J18027$-$1455 (central panel) and IGR
J21247+5058 (lower panel) acquired with the Cassini telescope at Loiano.
The spectra, in the 3900-8500 \AA~range, are smoothed with a Gaussian
filter with $\sigma$=4 \AA~(i.e. comparable with the spectral dispersion).
The main spectral features are labeled. The symbol $\oplus$ indicates
atmospheric telluric features. From Masetti et al. (2004a).}
\end{figure}

\vspace{0.1cm}
\noindent
{\bf IGR J17303$-$0601}. This source positionally coincides with an X--ray
object detected earlier by {\it ROSAT}, {\it HEAO-A1} and {\it RXTE} from
0.1 to 30 keV.  Within the small {\it ROSAT} error box (7$''$; Voges et
al. 1999) our $R$-band imaging showed at least 5 objects. The spectrum of
the brighter object in the error box (Fig. 1, top panel) revealed the
presence of Balmer and He~{\sc ii} emissions and interstellar absorption
features superimposed onto a reddened continuum. We therefore regarded the
identification of this source as the optical counterpart to IGR
J170303$-$0601 as secure. All spectral lines are at redshift zero, 
indicating that this object belongs to the Galaxy. Besides, the presence 
of the He~{\sc ii} strongly indicates undergoing mass accretion onto a 
compact star (e.g. van Paradijs \& McClintock 1995). We thus concluded 
that this source is very likely an X--ray binary system.

Using the strength of the Balmer emission lines we inferred a reddening of
$E(B-V)$ = 0.45 mag for the source. From this, and assuming a distance $d
\sim 8$ kpc, we obtained an absolute optical $R$-band magnitude $M_R \sim
0$ for the object, which is typical of persistent LMXBs (van Paradijs \&
McClintock 1995). The {\it INTEGRAL} X--ray data (Bird et al. 2004)
support this interpretation: at this distance, the 20--40 keV luminosity
of the source is 1.8$\times$10$^{35}$ erg s$^{-1}$. This also is typical
of persistent LMXBs in the soft state (e.g., Masetti et al. 2004b).

Recently, G\"ansicke et al. (2005) questioned the LMXB identification for
IGR J17303$-$0601 on the (admittedly sole) basis of the detection of a
128-second optical periodicity from this source, which they attribute to
the spin period of the accreting object. They conclude that, given this
spin duration, the compact source is a white dwarf and the system is
actually an Intermediate Polar CV similar to AE Aqr. However, one should
note that the NS hosted in the LMXB GX 1+4 has a comparable spin period
(134 s; e.g., Makishima et al. 1988). Moreover, using the information in
G\"ansicke et al. (2005), one derives, in the CV hypothesis, a distance of
$\sim$3 kpc to the object: this implies an X--ray luminosity of
$\sim$10$^{34}$ erg s$^{-1}$, which is four orders of magnitude larger
than that of AE Aqr (Mukai 2005). Thus the CV identification for this
source is not certain and it is moreover problematic when compared to the 
LMXB interpretation.

\vspace{0.1cm}
\noindent
{\bf IGR J18027$-$1455}. Inside the 2$'$ ISGRI error box of this source an
X--ray and radio object is found (Combi et al. 2005); it also has NIR and
optical ($R \sim 15$ mag) counterparts. The non-pointlike appearance of
the optical source suggested that it has extragalactic origin. Indeed, its
optical spectrum (Fig. 1, central panel) showed a faint and reddened
continuum dominated by a strong emission around 6800 \AA~which we
identified with the line complex composed of H$_\alpha$ plus [N~{\sc ii}]
at $z$ = 0.035 $\pm$ 0.001. Fainter and narrower emissions were also found
at wavelengths consistent with this redshift.  The spectrum allowed us to
identify this source as a Type 1 Seyfert galaxy.

Assuming a cosmology with $H_0$ = 65 km s$^{-1}$ Mpc$^{-1}$,
$\Omega_\Lambda$ = 0.7 and $\Omega_{\rm m}$ = 0.3, this redshift implies a
distance of 166 Mpc, X--ray luminosities of 3$\times$10$^{42}$ erg
s$^{-1}$ and 1.7$\times$10$^{44}$ erg s$^{-1}$ in the 0.1--2.4 keV and in
the 20--100 keV bands, respectively, and an absolute optical $B$-band
magnitude $M_B \sim -22$. These values place this source among the
brightest Type 1 Seyfert galaxies detected so far (Malizia et al. 1999;
V\'eron-Cetty \& Ver\'on 2003).

\vspace{0.1cm}
\noindent
{\bf IGR J21247+5058}. This source was associated by Combi et al. (2005)  
with the radio source 4C50.55, which shows a morphology typical of a radio
galaxy (Mantovani et al. 1982). The optical counterpart has magnitude $R
\sim 15.5$. Its optical spectrum (Fig. 1, bottom panel) has a puzzling
appearance. It shows a smooth continuum, typical of a late F- or early
G-type star in the Galaxy. However, superimposed to this stellar-like
continuum, a broad emission bump around 6700 \AA~is apparent, topped by a
narrow emission. By identifying the latter as H$_\alpha$, we obtained a
redshift $z$ = 0.020 $\pm$ 0.001.

The hypothesis of a chance alignment between a Galactic F-type star and a 
background radio galaxy at $z$ = 0.02 was thus suggested. In this 
occurrence, assuming the same cosmology as for IGR J18027$-$1455, the 
distance to this galaxy is 94 Mpc, and the extension of the radio lobes is 
$\sim$200 kpc; the 20--100 keV luminosity would be 1.4$\times$10$^{44}$ 
erg s$^{-1}$, locating this source also at the bright end of the AGN 
luminosity distribution. 
One should note that, although Mantovani et al. (1982) suggested that 
optical spectroscopy would be able to disentangle the nature of this 
object, no observations of this kind have been reported in more than 
20 years.

\smallskip

Therefore, our first campaign on the identification of the nature of
unknown {\it INTEGRAL} sources through optical spectroscopy gave
encouraging results, and indicates that the approach we used can indeed be
remarkably successful in this identification task.

\vspace{-.4cm}
\section{Future prospects}

\vspace{-.3cm}
Of course, {\it INTEGRAL} keeps monitoring the hard X--ray sky: the
2$^{\rm nd}$ IBIS survey is about to be issued (Bird et al. 2006), and new
sources are collected in the continuously updated {\it INTEGRAL} sources
inventory of Rodriguez (2005). Up to the time of this meeting (late May
2005), among this new set of {\it INTEGRAL} sources, some have been
optically identified as a LMXB containing a millisecond X--ray pulsar 
(Roelofs et al. 2004), three Be/X MXRBs (Reig et al. 2005; Halpern \& 
Gotthelf 2004; Torrej\'on \& Negueruela 2004), three supergiant MXRBs 
(Negueruela et al. 2005 and references therein) and a Type 1 AGN 
(Torres et al. 2004).

However, there is still a number of {\it INTEGRAL} sources which lack a 
firm optical/NIR identification of their nature: indeed, we have 13 
transient or persistent hard X--ray sources with (basically 
{\it INTEGRAL}) error boxes which are still too large to allow sensible 
optical/NIR counterpart searches; moreover, there are also 23 new sources 
(to which 12 more, belonging to the 1$^{\rm st}$ IBIS survey, should be 
added) which have soft X--ray or radio error boxes of size $\sim$10$''$ 
but non-univocal optical/NIR localization.

Therefore, for most of the above objects, very precise localizations with 
arcsecond or subarcsecond positional accuracy are needed to remove the 
ambiguity in the optical/NIR identification of the counterpart, especially 
in crowded fields. At present, this task can be achieved with the use of 
the satellites {\it Chandra}, {\it XMM-Newton} and {\it Swift} in X--rays 
and with facilities sensitive at radio wavelengths, in particular the VLA.

Of course, our optical follow-up activities are ongoing: we are about to
spectroscopically study northern and southern objects thanks to optical
observing programmes approved at ESO (Chile), SAAO (South Africa), WHT
(Canary Islands, Spain) and Loiano (Italy). These will secure new
identifications of the nature of more {\it INTEGRAL} sources.

However, besides this identification task, more work is expected in the 
future on these sources. In the following I briefly outline the main 
objectives which should be pursued:

\begin{itemize}

\item
\vspace{-.3cm}
cross-correlations with available catalogues at longer wavelengths is 
needed as soon as new {\it INTEGRAL} sources are discovered;

\item
\vspace{-.3cm}
a campaign of accurate X--ray astrometry of loose localizations with 
{\it Chandra}, {\it XMM-Newton} and {\it Swift} should be performed in 
order to pinpoint the soft X--ray counterparts of {\it INTEGRAL} sources;

\item
\vspace{-.3cm}
the study of the multiwavelength properties of the {\it INTEGRAL} sources, 
both considering each single case and grouping them into (sub)classes;

\item
\vspace{-.3cm}
the analysis of `Norma Arm sources': is it, as suggested by some 
authors (e.g., Filliatre \& Chaty 2004), a new class of objects? If so, it 
will disclose brand new information on heavily absorbed Galactic MXRBs;

\item          
\vspace{-.3cm}
the study of statistical correlations between INTEGRAL and surveys at 
wavelengths longer than soft X--rays;

\item
\vspace{-.3cm}
a statistical analysis of the nature of the objects belonging to the IBIS 
survey;

\item
\vspace{-.3cm}
in-depth observations to obtain precise physical parameters of the 
identified {\it INTEGRAL} objects.

\end{itemize}

\vspace{-.3cm}
To conclude, I stress the fact that, in the era of large observatories,
high-quality science on up-to-date astrophysical topics, such as the hunt
for the nature of {\it INTEGRAL} sources, can be achieved using small- and
medium-sized telescopes, as shown in Masetti et al. (2004a).

\begin{acknowledgements}
I would like to thank Franco Giovannelli for having given me the
opportunity of presenting this review at this meeting and the LOC for the
warm hospitality and the pleasant stay in Vulcano. 
\end{acknowledgements}

\medskip
\noindent
{\bf DISCUSSION}
\medskip

\noindent
{\bf NINO PANAGIA:} Can you comment on the fact that the spectrum of IGR 
J18027$-$1455 appears to be heavily reddened? Is this due to absorption 
intrinsic to the object?
\medskip

\noindent
{\bf NICOLA MASETTI:} Most likely the observed reddening is mainly
produced by foreground Galactic absorption, which is quite high along this
line of sight as it passes right through the Galactic Plane (the Galactic
latitude of the object is $b$ = +3$\fdg$7). Indeed, according to Schlegel
et al. (1998), a color excess $E(B-V)$ = 1.26 mag (which implies a
$V$-band extinction of nearly 4 magnitudes), induced by the Galactic dust,
is present in the direction of this source. 
\medskip

\noindent
{\bf ANATOLY IYUDIN:} Can you tell me how the distance to IGR
J16138$-$4848 was determined? As far as I know, its position is not
exactly in the middle of the Norma Arm, so the absorption along the line
of sight of this source should not be extremely high. 
\medskip

\noindent
{\bf NICOLA MASETTI:} Details on the determination of the distance to IGR 
J16138$-$4848 can be found in Filliatre \& Chaty (2004); however, let me 
remark that, as stressed by these authors, most of this absorption is 
local to the binary system and not due to the interstellar medium along 
the line of sight.

\end{document}